\documentclass[vecphys]{svmult}

\usepackage{makeidx}         % allows index generation
\usepackage{graphicx}        % standard LaTeX graphics tool
\usepackage{multicol}
\usepackage{color}        % used for the two-column index
\usepackage{url}             % allows inclusion of URLs
\usepackage[bottom]{footmisc}% places footnotes at page bottom
\usepackage{natbib}
\newcommand{\mnras}{MNRAS}
\newcommand{\apj}{ApJ}
\newcommand{\aap}{A\&A}
\newcommand{\pasj}{Publ.\ Astron.\ Soc.\ Japan}
\newcommand{\apss}{Astrophys.\ Space Sci.}
\newcommand{\araa}{Ann.\ Rev.\ Astron.\ Astrophys.}
\newcommand{\lesssim}{\mathrel{\hbox{\rlap{\hbox{\lower4pt\hbox{$\sim$}}}\hbox{$<$}}}}
\newcommand{\grsim}{\mathrel{\hbox{\rlap{\hbox{\lower4pt\hbox{$\sim$}}}\hbox{$>$}}}}
\newcommand{\eqb}{\begin{eqnarray}}
\newcommand{\eqe}{\end{eqnarray}}
\newcommand{\diff}{{\rm d}}
\newcommand{\rlight}{r_{\rm L}}
\newcommand{\degr}{^{\rm o}}

\makeindex             % used for the subject index
\begin{document}

\title*{The theory of pulsar winds and nebul\ae }


\author{J.G. Kirk\inst{1}\and Y. Lyubarsky\inst{2}\and J. P\'etri\inst{1}}


\institute{%
Max-Planck-Insititut f\"ur Kernphysik, Postfach 10 39 80, 69029
Heidelberg, Germany \texttt{John.Kirk@mpi-hd.mpg.de},
\texttt{Jerome.Petri@mpi-hd.mpg.de} \and Department of Physics,
Ben Gurion University, P.O.Box 653, Beer-Sheva 84105, Israel
\texttt{lyub@bgu.ac.il} }

\maketitle

\begin{abstract}
We review current theoretical ideas on pulsar winds and their surrounding 
nebul\ae. Relativistic MHD models of the wind of the aligned rotator, 
and of the striped wind, together with models of magnetic dissipation are 
discussed. It is shown that the observational signature of 
this dissipation is likely to be point-like, rather than extended,
and that pulsed emission may be produced. The possible pulse shapes and
polarisation properties are described. Particle acceleration at the
termination shock of the wind is discussed, and it is argued that two distinct
mechanisms must be operating, with the first-order Fermi mechanism producing 
the high-energy electrons (above $1\,$TeV) and either magnetic annihilation
or resonant absorption of ion cyclotron waves responsible for the $100\,$MeV
to $1\,$TeV electrons. Finally, MHD models of the morphology of the 
nebula are discussed and compared with observation. 
\end{abstract}

\section{Introduction}
\label{intro}
The theory of pulsar winds and the nebul{\ae} they energise is currently
celebrating its golden jubilee. Ten years before the discovery of
pulsars it was already apparent that the
magnetic field and relativistic particles 
that produce the radiation of the Crab
Nebula must have their origin in a central stellar object
\citep{piddington57}. Today, about 50 similarly powered 
objects are known, and some of them,
like the Crab, are detected and even resolved 
at all accessible photon frequencies, from the
radio to TeV gamma-rays. 
The rotation of the central neutron star
\citep{pacini67} is now 
universally accepted as the energy source fuelling these objects, but the
details of the coupling mechanism 
are still unclear. 
In this article we review  
current theoretical ideas on this subject and their relationship to
observations. We concentrate on the
magnetohydrodynamic description of the relativistic outflow driven by the
pulsar and on the bubble it inflates in the surrounding medium.  

The discussion is organised as follows: in 
section~\ref{magnetosphere}
we consider the region between the surface of the neutron star and the {\em
light cylinder} a surface of cylindrical radius $\rlight=cP/(2\pi)$, where $P$
is the pulsar period. The speed of an object that corotates with the star
becomes luminal on this surface, and the wavelength of the radiation that
would be emitted by the pulsar in vacuum is $2\pi\rlight$. 
In the terminology of radiating systems, the region within the light 
cylinder is, therefore, the \lq\lq near zone\rq\rq, where
the fields
can be approximated as being in rigid corotation.  
Conventionally, this region is called
the pulsar {\em magnetosphere}. It is thought to be the site of copious pair
creation, and, in most theories, is the region in which the pulsed radiation
itself is emitted. 

Electric and magnetic fields dominate the dynamics in the near
zone inside the light cylinder.  However, this is also true in at least part
of the \lq\lq far zone\rq\rq, 
that lies well outside the light cylinder --- a region that is
not conventionally thought of as the pulsar magnetosphere. Here, we follow
conventional terminology,
and call this the {\em wind zone}\footnote{%
Note that we designate the entire far zone outside the light cylinder
as the  {\em pulsar wind},
although this term is really only appropriate for the supersonic part of the
flow.};
it extends up to
the {\em termination shock}, where the ordered, relativistic flow of particles
and fields is randomised.
Electromagnetic fields dominate the inner part
of the wind zone, where the plasma flows subsonically with respect to the fast
magnetosonic mode. As we describe in section \ref{aligned}, 
force-free solutions to the aligned rotator problem show that 
the plasma accelerates up to the critical
point where its velocity equals the fast magnetosonic speed, 
and, presumably, passes 
through it into a supersonic
domain. In this part
of the flow, the dynamics is strongly influenced by the particles, although
most of the energy flux continues to be carried by the fields.
The oblique rotator is a more suitable model for a pulsar, and the key
difference it introduces in the wind zone 
is wave-like structure on the very small 
length scale $\rlight$. 
We discuss the structure of such a {\em striped wind}
in section~\ref{stripedwind}, and consider
the possibility that 
dissipative processes play a role in
converting Poynting flux into kinetic energy.
In both the sub and supersonic parts, the 
pulsar wind is highly relativistic. If it radiates at all, its emission
is beamed 
predominantly in the radial direction. This simple kinematic effect has an
important influence on the radiative signature of this region, as we discuss
in section \ref{observability}.

Observationally, the termination shock can be identified as the outer boundary
of an underluminous region that lies at the centre of the diffuse synchrotron
emission of the nebula. The nebula itself, therefore, lies between the
termination shock and an outer boundary, where the relativistic particles
produced by the pulsar 
impinge on either the debris of its natal supernova explosion, or on the
interstellar medium. 
These particles are 
most plausibly accelerated at the
termination shock, and we consider the 
acceleration processes operating here
in section~\ref{terminationshock}.
Finally, in section~\ref{nebula}, we
discuss models of the nebula and its emission.

\section{The magnetosphere}
\label{magnetosphere}

An enormous electric field is induced by the
rotation of a magnetised neutron star. This field tears particles from the
stellar surface and accelerates them up to high energies. 
As a result, these
\lq\lq primary\rq\rq\ particles initiate an electron-positron cascade, which,
according to conventional wisdom, 
populates the entire magnetosphere with plasma. In the aligned case, 
solutions have been found for the region well within the light cylinder
in which
this plasma is confined to domes
above the poles and a differentially rotating equatorial disk
\citep{krause-polstorffmichel85,petrietal02}. If, on the other hand,
the magnetosphere is filled with plasma, 
the strength of the magnetic field is sufficient to
ensure that the plasma corotates 
\citep[for recent reviews, see][]{petri06,hardinglai06}.
At some point near the light cylinder, corotation must cease, 
and the particles escape, carrying away magnetic flux and 
energy in the form of an ultrarelativistic, magnetised wind. 

If one traces the escaping (\lq\lq open\rq\rq)
magnetic field lines back to the stellar surface, 
they define areas which, in a dipole geometry, lie close to the magnetic
axis and are called \lq\lq polar caps\rq\rq.  
To maintain a steady state,
plasma must be continuously generated on the open field lines 
in these polar regions. It then streams along
them with relativistic velocity and
eventually escapes through the light cylinder. 
It is usually assumed that the 
mechanism responsible for creating the pulsed
radiation is somehow associated with the materialisation of 
these plasma streams, which presumably takes place
at either an \lq\lq inner gap\rq\rq\ or 
\lq\lq outer gap\rq\rq, \citep[e.g.,][]{harding05}. However, this is
not necessarily true (see section~\ref{observability}), and,
at least for the optical pulses, there are indications
that the site of pulse production could lie outside the light 
cylinder \citep{kirketal02,dyksetal04,petrikirk05}. 

The rate at which pairs escape is conveniently measured in terms
of the the pair multiplicity, $\kappa$, which is 
the number of pairs produced by a single
primary particle that emerges from one of the polar caps. 
The primary beam consists of particles of a single charge, and its density
is expected to be close to the
Goldreich-Julian density, defined as that 
required to screen the induced electric field, namely
\eqb
n_{\rm GJ}&\equiv &\left|(\mathbf{\Omega\cdot
B})/\left(2\pi ec\right)\right|
\label{gjdensity}
\eqe
\citep[e.g.,][]{davis47,honesbergeson65,michel91} where 
$\mathbf{\Omega}$ is the angular velocity of the star and 
$\mathbf{B}$ the magnetic field. In an 
aligned dipole geometry, 
the foot-points of those field lines
that cross the light cylinder
lie within polar caps of (cylindrical) radius 
$r_{\rm pc}\approx\sqrt{R_*^3/r_{\rm L}}$, 
where $R_*$ is the stellar radius, and this is a reasonable approximation 
also in the oblique case \citep[e.g.,][]{dyksetal04}.
Assuming the pairs stream away from the polar cap at relativistic speed, 
the total number ejected per unit time may therefore be
estimated as
\eqb
\dot N&=&\kappa\frac{\Omega^2B_*R_*^3}{ec}
\label{multiplicity}
\\
&=&2.7\times10^{30}\kappa \left({B_*\over10^{12}\,\textrm{G}}\right)
\left({P\over 1\,\textrm{s}}\right)^{-2}\textrm{s}^{-1}
\nonumber
\eqe
where $\mathbf{B}_*$ is the magnetic field at the polar cap. 

The actual value of the multiplicity, $\kappa$, is rather uncertain.  
Theoretical models 
\citep{hibschmanarons01a,hibschmanarons01b}, give values of $\kappa$
ranging from a few to thousands, but some 
observations suggest that the real value might be
substantially higher 
\citep[][see section~\ref{nebula}]{gallanttuffs02,bietenholzetal04}. 
If $\mathbf{\Omega\cdot B}_*<0$, the primary particles are positively charged
ions. Because these particles do not breed in cascades, they make up a
fraction of at most $1/\kappa$ by number of the wind particles. Nevertheless,
if $\kappa$ is indeed as low as predicted theoretically, the energy flux
carried by ions could be as large as that carried by pairs. 

\section{The wind of an aligned rotator}
\label{aligned}
Pairs are produced with an energy that ranges from dozens to hundreds of 
MeV, and the total energy density of the produced plasma remains
small compared to the  magnetic energy density. Nevertheless, 
the presence of the plasma is crucially important,
because the electromagnetic structure in the far zone
is strongly influenced by the currents it carries. 
The overall structure may be pictured by taking into account that the
magnetic field is frozen into the plasma. 
As the plasma cannot rotate with superluminal velocity, 
the magnetic field lines beyond the light cylinder 
are wrapped backwards
with respect to the rotation of the neutron star. 
As a result of this,
even an axisymmetric rotator 
loses energy by
driving a plasma wind, provided its inner zone is filled with 
plasma. In contrast, an aligned magnetic dipole rotating in
vacuum does not lose energy. 

In the case of an aligned rotator 
(with parallel magnetic dipole and rotation axes), the entire 
system is axisymmetric. 
Essentially, the rotational energy of the neutron star is spent in the 
generation of azimuthal magnetic field in the wind. 
This can be seen by noting that the electric
field ($\mathbf{E}'$) in the proper frame of the flow  vanishes 
because the
plasma conductivity is infinite, which implies
\begin{equation}
\mathbf{E}+(1/c)\mathbf{v\times B}=0.
\label{idealmhd}
\end{equation}
In a radial, relativistic wind, the poloidal component of 
$\mathbf{B}$ is also
radial, so that the magnitude of the electric field is close to that of the
azimuthal magnetic field, which then determines the radial component of the
Poynting flux: $P_R\approx cB_\varphi^2$.
 
In a steady, axisymmetric solution, the displacement current vanishes, 
and the azimuthal magnetic field is generated by poloidal currents
that flow either into or out of the polar caps of the 
star, depending on
the sign of $\mathbf{\Omega}\cdot\mathbf{B}_*$. 
The circuit is closed in a current 
carrying surface --- a 
\lq\lq current sheet\rq\rq. Well outside the light cylinder, this sheet 
lies in the equatorial plane and separates the field lines that originate 
from the 
two magnetic poles. Inside the light cylinder, the 
current flows in the surface that encloses the volume containing 
closed field lines.

At the light cylinder, the azimuthal and poloidal fields are comparable. In
the wind zone, conservation of magnetic flux in a diverging flow
implies that the poloidal
field decreases faster than the azimuthal field, 
which is proportional to $1/r$, where $r$ is the cylindrical radius.
Thus, the field in the far zone of the wind may be considered as 
purely azimuthal; even though each
magnetic field line is a spiral anchored on the surface of the star, the spiral
becomes so tightly wound in the far zone that it can be approximated locally
as separate coaxial magnetic loops moving together with the flow.

The relative strength of magnetic field and particles is
an important characteristic of the wind. This is best defined, 
in an ideal MHD description, as the   
ratio in the proper frame (where $\mathbf{E}'=0$) of the magnetic to
particle enthalpy densities:
\begin{equation}
\sigma\equiv\frac{{B'}^2}{4\pi w c^2}
\label{sigmaproper}
\end{equation}
where $w$ is the proper enthalpy density of the plasma and
$\mathbf{B}'$ is the magnetic field in the proper frame. 
If, as we assume, the plasma is cold, 
the enthalpy 
density  is simply the rest mass energy density: $w=m_{\rm e}c^2 n/\Gamma$,
where $n$ is the number density in the 
lab.\ frame, (in which the centre of mass of the neutron star is at rest)
and $\Gamma$ is the Lorentz factor of the wind in this frame.
Then, since the velocity of the wind is 
perpendicular to $\mathbf{B}'=\mathbf{B}/\Gamma$, 
$\sigma$ equals the ratio of the energy carried by Poynting flux to that
carried by particles:
\eqb 
\sigma&=&B^2/4\pi m_{\rm e}c^2n\Gamma
\label{sigma}
\eqe

According to all available models of the multiplicity, 
$\sigma\gg 1$ close to the light cylinder, so that the wind is 
Poynting dominated at that point. This means that the fast 
magnetosonic speed $v_{\rm fms}$ is very close to $c$:
\eqb
v_{\rm fms}&=&\frac{cB'}{\sqrt{4\pi w +B'^2}}
\nonumber\\
&\approx&
\sqrt{\frac{\sigma}{1+\sigma}}
\eqe
\citep[e.g.,][]{kirkduffy99} and the corresponding Lorentz factor
is $\Gamma_{\rm fms}=\sqrt{\sigma}$.
In the cold wind, $\sigma$ is related to $\Gamma$ via Eq.~(\ref{sigma}), so 
that the fast magnetosonic point is located 
at the point where 
\eqb
\Gamma&=&\Gamma_{\rm fms}\,=\,
\left(\frac{B^2}{4\pi m_{\rm e}c^2n}\right)^{1/3}
\\
&\equiv&\mu_{\rm M}^{1/3}
\eqe
where $\mu_{\rm M}$ is the magnetisation parameter introduced by
\citet{michel69}.
The propagation of the fast magnetosonic wave is a result of 
the interplay between magnetic tension and plasma
inertia. 
If, as seems likely, the flow emerges through the light cylinder subsonically,
i.e., with $\Gamma\ll\sqrt{\sigma}$, the dynamics simplify
significantly, because the inertia terms in the equation of motion are
unimportant.  
Since gravity and gas pressure are also unimportant in the pulsar case,
this \lq\lq force-free\rq\rq approximation implies that 
the Lorentz force is exactly cancelled by the 
electric force: 
\eqb
\rho_e\mathbf{E}+(1/c)\mathbf{j\times B}&=&0
\label{forcefree}
\eqe
The description of  the dynamics is complete when this 
equation is complemented by Maxwell's equations, connecting
the charge and current densities $\rho_e$ and $\mathbf{j}$ with the
fields $\mathbf{E}$ and $\mathbf{B}$, and by the ideal MHD condition
(\ref{idealmhd}).

The force-free MHD equations for the pulsar magnetosphere
are strongly nonlinear and 
must, in general, be solved numerically
\citep{contopoulosetal99,gruzinov05,komissarov06,mckinney06,timokhin06}. 
In particular,
the oblique rotator,
being a three-dimensional problem, has only recently been treated
\citep{spitkovsky06}. 
Nevertheless, 
an exact axisymmetric solution describing a
magnetosphere of a rotating star, namely that of the split
monopole, has been known for many years
\citep{michel73}. In this solution, magnetic field lines 
extend from the
origin to infinity in the upper hemisphere and converge from infinity 
to the
origin in the lower hemisphere. The hemispheres are separated by
an equatorial current sheet. The magnetic surfaces have the form of
coaxial cones whose vertices lie at the origin. Plasma flows
radially from the origin to infinity. Of course, the flow lines in
a realistic (dipole) magnetosphere cannot be radial everywhere;
there should be a zone of closed field lines inside the light
cylinder. However, 
\citet{ingraham73} and \citet{michel74} showed that, 
independently of the field structure
near the origin, the flow lines beyond the light cylinder become
asymptotically radial in the
force-free approximation, just as in the split monopole solution. 
This is in contrast with the situation in axisymmetric nonrelativistic 
MHD winds 
where hoop stresses collimate the flow along the 
rotation axis. The reason is that in the relativistic case,
the electric force 
compensates these stresses almost exactly.

In axisymmetric MHD, the flow lines lie in magnetic surfaces and the
electric field is perpendicular to these surfaces, which are, therefore,
equipotentials. The force-free condition (\ref{forcefree}) 
implies that the currents also flow along  
the magnetic surfaces. Taking
this into account, one can find the asymptotic behaviour of the
basic quantities in the radial wind:
\eqb
\begin{array}{rcl@{\qquad}rcl}
B_{\varphi}&\propto&{r_{\rm L}}/{R}&
E&\propto&{r_{\rm L}}/{R}
\\
B_R&\propto&{r_{\rm L}^2}/{R^2}&\quad n&\propto&
{r_{\rm L}^2}/{R^2}
\end{array}&&
\label{asymptoticsol}
\eqe
where $R$ is the radius in spherical polar coordinates.
It follows from the higher order terms in the 
asymptotic solutions presented by \citet{ingraham73}
and \citet{michel74} that the difference 
between $E$ and $B$ decreases with radius and 
the flow velocity approaches 
closer and closer to $c$, i.e., it accelerates. 
\citet{buckley77} showed that, in the
force-free approximation, the Lorentz factor of the flow grows
linearly with the radius \citep[see also][]{contopouloskazanas02}. 
Eventually, the Lorentz factor becomes comparable to that of the fast
magnetosonic mode $\Gamma_{\rm fms}=\sqrt{\sigma}$. The plasma inertia becomes
important at this point, and the force-free approximation breaks down. 

If the pulsar wind is launched with $\sigma_0\gg
1$ and $\Gamma_0\ll\sqrt{\sigma_0}$, it is initially subsonic, 
the force-free approximation holds, and the flow accelerates.
As $\Gamma$ increases, $\sigma$ decreases, since, in
a radial flow with purely toroidal field, $B^2/n$ 
remains constant and $\sigma\propto
\Gamma^{-1}$ according to eq.~(\ref{sigma}). 
At the fast magnetosonic point, $\sigma=(\sigma_0\Gamma_0)^{2/3}\gg 1$, 
and the flow is still Poynting dominated. This is in contrast with the
nonrelativistic situation, 
where $v_{\rm fms}=B/\sqrt{4\pi\rho}$ so that
the energy flux carried by particles equals the Poynting flux at the 
fast magnetosonic point. 
The reason is that in the relativistic case, 
the electric and magnetic forces almost compensate each other, 
allowing inertial effects to come into play at an early stage.

Beyond the fast magnetosonic point 
the full relativistic MHD
equations must be solved. However, the flow remains
nearly radial even there 
\citep{tomimatsu94,beskinetal98,chiuehbegelman98,bogovalovtsinganos99,
lyubarskyeichler01}, because the additional inertial forces 
tend to resist collimation. Thus, if the flow is radial in
the force-free region, it remains radial further out. In fact, the
flow is practically ballistic, i.e., the plasma moves radially
with negligible acceleration, so that $\Gamma$ and, hence, $\sigma$ remain
constant. Thus, within the scope of ideal MHD,
the wind can be accelerated to at most Lorentz factors of a few
times $\Gamma_{\rm fms}$, at which point $\sigma$ is still large. 
Beyond this point, 
electromagnetic energy is not transferred to the
plasma.\footnote{%
This does not mean that
$\sigma$ remains large in {\it any} Poynting-dominated flow.
The relativistic MHD equations do not forbid acceleration
of the flow and conversion of Poynting flux into kinetic energy.
There exist many solutions
that demonstrate explicitly the reduction 
of Poynting flux to the equipartition
level or even below \citep{lietal92,vlahakis04,beskinnokhrina06}. 
However this can happen only if the flow lines are not radial. In
the above-mentioned solutions, this is achieved by a special
choice of the poloidal flux distribution such that the total
poloidal flux is infinite and the magnetic surfaces do not
converge to origin. (Note that although the poloidal field is small
beyond the light cylinder, the stress of this field may be not
negligible in some cases because the hoop stress is nearly
compensated by the electric force.) 
Whereas it is quite plausible that such 
solutions could be matched to boundary conditions above a 
disk, it is difficult to imagine that
they could be made compatible with a star (effectively
a point source) threaded by a finite magnetic flux. 
For this reason, we do not discuss them further here.}

\begin{figure}
\includegraphics[bb=129 263 476 540,width=\textwidth]{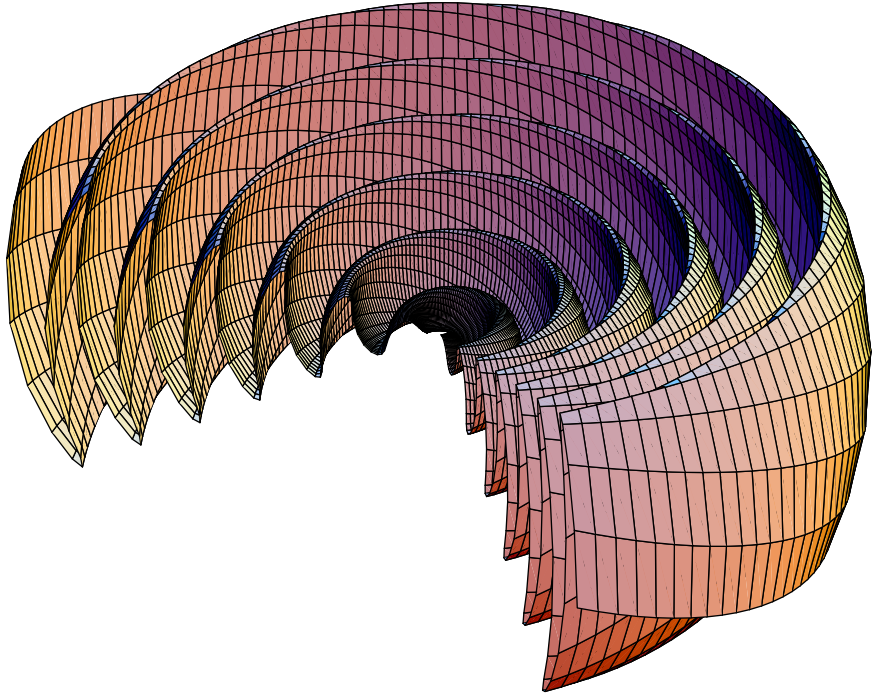}
\caption{%
\label{stripedwindfig}%
The {\em striped wind} \citep{michel71,coroniti90}. 
A snapshot is shown of the surface that is traced out by the inclined magnetic
equator when it rotates and 
is carried outwards by a radially propagating wind of constant velocity. 
This surface corresponds precisely
to the current sheet in the oblique split-monopole solution
\citep{bogovalov99}.  
}
\end{figure}

\section{The striped wind}
\label{stripedwind}
In the real world, pulsar winds are neither axisymmetric nor steady.  In fact,
the defining characteristic of a pulsar is that it is an oblique rotator, in
the sense that its magnetic and rotation axes are not parallel. The
time-varying electromagnetic fields excited by such an oblique rotator
propagate outwards in the form of electromagnetic waves. The wavelength of
these oscillations is at most $2\pi\rlight$. In the far zone, this is much
smaller than the radius. For this reason, dissipative processes, which 
operate only on short length scales, are likely to be much more important in
the wind of an oblique rotator than in a steady, axisymmetric wind.  
Therefore, the 
dissipation or damping of these 
waves could be an important mechanism of energy
transformation in a pulsar wind.

A rarefied magnetised plasma supports a variety of electromagnetic waves.
However, it seems reasonable to simplify the problem by assuming that only MHD
waves (i.e., those satisfying the ideal MHD condition (\ref{idealmhd}))
survive in the far zone.
The reason is that, according to the conventional picture, the plasma density
at the base of the pulsar wind (at $r\sim \rlight$) is sufficiently large that
low-frequency electromagnetic waves are heavily damped
\citep[e.g.,][]{asseoetal78,melatosmelrose96}.
In this case, it is once 
again useful to think of magnetic field lines as frozen
into the plasma flow. 
In the equatorial belt, the magnetic field at a
fixed radius alternates in direction at the frequency of
rotation, being connected to a different magnetic pole every
half-period. \citet{michel71} pointed out that the flow in this zone
should evolve into regions of cold, magnetically dominated plasma
separated by very narrow, hot current sheets. Such a structure 
can be thought of as containing 
four tangential discontinuities per wavelength,
each of which separates field lines from one magnetic pole from the thin,
bounding layer of hot plasma which constitutes the current sheet. The 
corresponding structure in hydrodynamics is called 
an entropy wave. It is simply a stationary
pattern of spatially varying temperature and
density that is in pressure equilibrium in a motionless plasma. 

The global picture of the resulting flow is shown in figure 
\ref{stripedwindfig}. As in the aligned rotator, the current sheet is a single
surface that separates magnetic field lines that are attached to 
opposite magnetic hemispheres on the stellar surface.  However, in the oblique
case a  pattern similar to that observed in the solar wind emerges:
the sheet develops corrugations, whose amplitude increases linearly with
distance from the star, as the radial wind draws out the flux lines. 
The current sheet now cuts the equatorial plane along twin spirals, that 
separate stripes of magnetic field of opposite polarity --- hence the name
{\em striped wind}. 
In the far zone, the distance between successive corrugations is small
compared to the radius of curvature of their surfaces, and the 
spiral in the equatorial plane becomes tightly wound. 
Within a belt around the equator, whose thickness depends on the angle between
the magnetic and rotation axes (see below), the flow locally
resembles a sequence of concentric, spherical current sheets.

This picture can be made more precise by noting that, in an ideal MHD solution,
the polarity of the field between two magnetic surfaces can be reversed without
affecting the structure of the solution, provided a current sheet is inserted
on the bounding surfaces. \citet{bogovalov99} applied this argument to the
split monopole solution by anchoring the inner edge of the 
current sheet to the obliquely rotating
magnetic equator on the stellar surface. This effectively transforms the
aligned split-monopole solution into one for an obliquely rotating split
monopole.  
The resulting picture coincides 
precisely with that illustrated in figure \ref{stripedwindfig}.
Recently, numerical solutions of the force-free equations have been
found that have a similar appearance
\citep{spitkovsky06}, although the dimensions of the 
calculational box extend as yet to only a 
few$\,\times\,\rlight$. 

At high latitudes, the magnetic field does not change sign,
and there are no current sheets embedded in the flow. Here, 
the magnetic oscillations can propagate as fast
magnetosonic waves; the generation 
of such waves by the rotating, slightly
nonaxisymmetric magnetosphere was considered by 
\citet{bogovalov01}. They could decay by nonlinear 
steepening leading to the formation of multiple shocks
 \citep{lyubarsky03a}. But 
this could release only a fraction of the Poynting flux into the plasma, 
since, at these latitudes, most of it is carried by the mean 
magnetic field.

The X-ray image of the inner Crab Nebula 
clearly
suggests that most of the energy is transported in the equatorial
belt of the pulsar wind
\citep{aschenbachbrinkmann75,brinkmannetal85,weisskopfetal00}. 
The split monopole solution 
indeed has a
pronounced maximum of the Poynting flux 
at the equator:
$\diff L/\diff\Omega\propto\sin^2\theta$, where $\diff L/\diff \Omega$ is the 
luminosity per solid angle interval and $\theta$ is the colatitude 
\citep{michel73}. In an equatorial belt, 
$\pi/2 - \zeta <\theta<\pi/2 + \zeta$, where $\zeta$ is the angle between the
magnetic and rotation axes, the energy is carried predominantly by 
alternating fields (the mean field of oblique rotator
vanishes in the equatorial plane). 
This means that most of the energy is transported in the
form of a striped wind; therefore the fate of the striped wind
is of special importance.

In an ideal MHD picture, the entropy wave that makes up the
striped wind propagates without damping, and the dynamics is the same as
in the case of the aligned rotator: in the supersonic region, the 
flow is essentially
ballistic and propagates radially at constant speed. However, as noticed by
\citet{usov75}, this cannot continue to arbitrarily large radius. The reason 
is that
the amplitude of the magnetic oscillations, which is proportional to 
the current
flowing in the sheets, decays only as $1/R$, whereas the number of particles
contained by the sheet decays as $1/R^2$. At some radius, therefore, 
there cease to be
enough particles to carry the required current. 
 
This problem can be avoided, or at least postponed, if the current sheet is
able to \lq\lq recruit\rq\rq\ additional charge carriers from the surrounding,
magnetised plasma. However, such a process corresponds to the annihilation of 
the magnetic flux that originally threaded the newly recruited charge
carriers. It hinges on the existence of an entropy creating dissipation
mechanism, and the rate at which it can proceed depends on the details of the
microphysics of this mechanism. 

A convenient analytical approach to this problem 
is to employ the small wavelength approximation. This was first done by 
 \citet{coroniti90}. He adopted an implicit model of the dissipation 
by assuming that it proceeded just fast 
enough to maintain the minimum required number of 
charge carriers. An equivalent formulation of 
this assumption is 
the requirement that the thickness of the current sheet should be 
equal to the gyro-radius of the sheet particles. 
The full set of relativistic 
MHD equations, complemented by this assumption about 
dissipation in the sheet, was solved by \citet{lyubarskykirk01} in the small
wavelength approximation. They found that dissipation causes the 
supersonic flow to
accelerate, thus effectively converting 
Poynting flux into kinetic energy. Unfortunately, an inescapable side effect
of this acceleration is the relativistic dilation of the dissipation time
scale. Taking account of this, one finds that the Lorentz factor of the flow
grows only slowly, according to $\Gamma\propto R^{1/2}$. 
Applying these results to the Crab Nebula, \citet{lyubarskykirk01} concluded
that this kind of dissipation could not convert a significant fraction of the 
wind luminosity into kinetic energy before the flow encountered the
termination shock. 

The implicit assumption about the dissipation rate in this calculation 
is clearly both fundamental and arbitrary. In an attempt to improve this
situation, \citet{kirkskjaeraasen03} compared the effects of three different
prescriptions for the dissipation rate. As well as the minimum rate 
used by \citet{coroniti90} and \citet{lyubarskykirk01}, they found solutions
corresponding to dissipation on the timescale of the growth of the
relativistic tearing mode, and to dissipation at 
the maximum plausible rate, governed by the 
transit time of sound waves across the sheet.  Applying these to the Crab 
Nebula, they concluded that conversion of the Poynting flux to kinetic energy
was indeed possible in the most favourable case, but only if the 
outflow carried substantially more pairs than predicted by the cascade models 
of \citet{hibschmanarons01a,hibschmanarons01b}.

Although the short-wavelength approximation enables one to 
find analytical solutions and sketch out possible scenarios
for the solution of the $\sigma$ problem, it does not necessarily  
follow that these scenarios are realised in any given pulsar. 
It could be, for example, that dissipation becomes
important even before the wind is accelerated to supersonic speed. 
The deposition of a substantial amount of energy into heat
in the wind zone is 
likely to result in an observable signature, as discussed in
section~\ref{observability}. 
Finally, if the analytical solutions indeed describe the wind accurately, it 
could be that relatively little Poynting flux is converted into kinetic energy
before the termination shock is reached. The observed morphology of the  
Crab Nebula could, nevertheless, be recovered if the conversion were to take
place instead in the termination shock itself \citep{lyubarsky03b}. 
According to the results of one-dimensional
particle-in-cell simulations \citep{lyubarsky05,petrilyubarsky07}, 
this appears
plausible, and may open up a way to understand the particle acceleration
process operating at this shock front (see section \ref{terminationshock}).

\section{Observability of the wind}
\label{observability}

\subsection{Point-like appearance}
\label{pointlike}
Whereas the termination shock and nebula are visible in the X-ray
\citep{weisskopfetal00} and optical \citep{hesteretal95,hesteretal02} images
of the Crab Nebula, the wind zone they enclose appears to be dark, as was
noticed in early optical images \citep{scargle69}. 
The standard explanation of this phenomenon is that 
the MHD wind is expected to be cold. In the comoving frame, a volume
element in a ballistic wind would expand by a factor of $10^{18}$ between the
light cylinder and the termination shock of the Crab, so that any random
motion should quickly disappear. Cold, ordered motion of a highly conducting
plasma, does not, however, produce synchrotron radiation. According to
the ideal MHD condition, eq.~(\ref{idealmhd}), the acceleration of each
particle vanishes, so that all trajectories are rectilinear. I.e., 
in the comoving frame,
all particles are stationary. The emissivity for bremsstrahlung and for other
\lq\lq thermal\rq\rq\ processes also vanishes in this case. 
The only possibility of
producing radiation in this case is by coupling the bulk motion to the photon
field. Inverse Compton scattering of photons of the cosmic microwave
background or other target fields could do this, but would give rise to
gamma-rays rather than X-rays or optical photons
(see section~\ref{icscattering}). Unfortunately, the resolution of 
current gamma-ray 
detectors does not enable such photons to be distinguished 
from the nebular emission. 

However,
even if the plasma somehow remains hot, perhaps because of internal
dissipation, the emission should be strongly beamed into a cone of
(half)-opening angle $1/\Gamma$ in the radial direction.  One
might naively expect that the image of the wind on the sky should not exceed
an angular size of $1/\Gamma$.  In fact, the maximum possible size of the
image is much smaller, and depends on the radial dependence of the emissivity,
as well as on the radial dependence of $\Gamma$.
To see this, consider a simple model in which the emissivity of the wind is
such that emission from all radii $R$ is visible, provided that the radius
vector $\mathbf{R}$ makes an angle of less than $1/\Gamma(R)$ with the line of
sight, and provided that $R<R_{\rm T}$, where $R_{\rm T}$ is the radius of the
termination shock, here assumed, for simplicity, to be spherical in shape. 
\begin{figure}[htbp]
  \centering
  \scalebox{0.5}{
\input{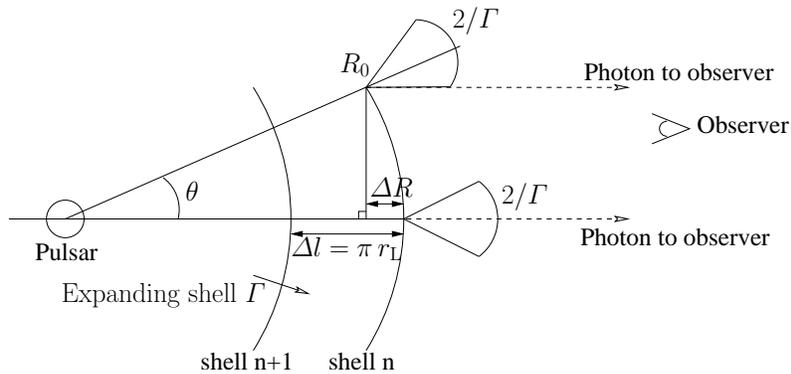}
}
  \caption{Spherically symmetric shells expanding 
    with relativistic Lorentz factor $\Gamma$ and emitting within a
    cone of opening angle $2/\Gamma$ when crossing the surface $R_0$.}
  \label{fig:Shell}
\end{figure}
The solutions found in the short-wavelength approximation
(section~\ref{stripedwind}) suggest the Lorentz factor can be parameterised as 
\eqb
\Gamma(R)&=&\Gamma_{\rm T} \, \left( \frac{R}{R_{\rm T}} \right)^q
\label{eq:LorentzWind}
\eqe
where $\Gamma_{\rm T}$ is the Lorentz factor at the termination shock
and $q$ ($\ge 0$) describes the radial acceleration. 
Making the reasonable assumption that the distance $D$ to the pulsar is large,
specifically, $D\gg \Gamma_{\rm T}R_{\rm T}$, it follows that 
radiation is emitted in the
  direction of the observer provided $\sin\theta\le1/\Gamma(R)$, 
and $R\le R_{\rm T}$. 
This defines a surface that limits the visible part of the wind
to:
\eqb
R&\le&R_{\rm T}\textrm{Min}\left[1,
1/\left(\Gamma_{\rm T}\sin\theta\right)^{1/q}\right] 
\eqe
On the plane of the sky, this surface appears to have an angular 
diameter
$\alpha=\alpha_{\rm T}R\sin\theta/R_{\rm T}$ 
where $\alpha_{\rm T}$ is the angular diameter of 
the termination shock. Hence, the diameter of 
the visible part of the wind is
\eqb
\alpha&\le&\frac{\alpha_{\rm T}}{\Gamma_{\rm T}}
\textrm{Min}\left[\Gamma_{\rm T}\sin\theta,\left(\Gamma_{\rm
    T}\sin\theta\right)^{1-1/q}\right]
\eqe
For $q<1$, such as is found for all the dissipation models tested by
\citet{kirkskjaeraasen03}, the maximum 
angular diameter of the wind emission is reached at the termination shock:
\eqb
\alpha_{\rm max}&=&\frac{\alpha_{\rm T}}{\Gamma_{\rm T}}
\label{anglimit}
\eqe
whereas, for more rapidly accelerating flows ($q>1$), the maximum diameter is 
determined at the launching point. 
 
Optical observations reveal a point-like source, of angular size less than
$0.1\,$arcsec at the position of the pulsar, 
whereas the termination shock has an
angular diameter of approximately $12\,$arcsec. 
If there is substantial dissipation
within the wind, then these observations establish a minimum Lorentz factor
that is required in order that the wind should still appear point-like. For
$q<1$, this is simply 
\eqb 
\Gamma_{\rm T}&>&\frac{\alpha_{\rm T}}{\alpha_{\rm max}} 
\nonumber\\ 
&\grsim& 100 
\eqe 
This is a modest requirement when compared
to most estimates, which lie in the range from $\Gamma\sim10^5$ to $10^6$. In
the case of more rapid acceleration $q>1$, the constraint is even less
stringent. For a launching point $R_0$ it reads $\Gamma\grsim24 R_0/R_{\rm
T}\ll 100$. Thus, one concludes that the appearance of the optical and
X-ray nebula, with
its central point-like source and dark region within the termination shock,
does not rule out the possibility that a substantial fraction of the pulsar
luminosity is dissipated and radiated away in the wind zone, provided only
that the flow speed remains high. This same argument 
applies also to those parts of the nebula that remain in 
relativistic, approximately radial motion. Such regions 
are indeed found in two-dimensional MHD models
(see section~\ref{nebula}) and are expected to produce almost
point-like images close to the pulsar.

\subsection{Inverse Compton scattering}
\label{icscattering}

As mentioned above, the wind, though cold, could in principle be detected
through the emission produced by inverse Compton (IC) scattering of an
external, soft photon field \citep{ballkirk00,bogovalovaharonian00}.
There are several possible external origins 
of the target photon population: the cosmic microwave background
radiation, the synchrotron radiation of the nebula (dominant for the Crab 
from radio
wavelengths to X-rays), the thermal far-infrared radiation (probably
associated with dust), from the surroundings of the nebula
\citep{atoyanaharonian96} and, in the case of a pulsar in a binary system, 
photons from the companion star \citep{balldodd01}.  
For a wind of constant Lorentz factor, the
emission is very strongly peaked in frequency space. Even for an accelerating
wind, an almost monochromatic line should appear in the gamma-ray range.
The integrated (over frequency) luminosity of the scattered
photons is well below the sensitivity of current gamma-ray telescopes ---  
only if the peculiarities of the predicted spectral distribution 
could be exploited might it be possible to extract a signal.
 
Another possibility is an internal source of target photons, i.e., photons
from the pulsar itself. These photons suffer the obvious disadvantage that
they propagate radially outwards together with the wind. Scattering events are,
therefore, likely to be almost forward in direction, which implies relatively
small energy gain. However, if, by some mechanism, 
the wind becomes kinetic energy dominated quite close to the
light cylinder, the angular momentum it must carry ensures that its velocity
vector makes a substantial angle with the radius vector. Photons
from the stellar surface could then be significantly boosted in energy. 
\citet{bogovalovaharonian00} investigated this possibility for the Crab,
taking into account as target photons 
not only the thermal X-rays from the surface of this
young pulsar, but also the nonthermal, pulsed emission (assuming it too comes
from close to the stellar surface).  
Because the predicted
high-energy flux depends sensitively on the location of the transition from a
magnetically to a kinetically dominated wind, the observed gamma-ray flux
(which probably originates from the nebula) puts
a lower limit on this conversion radius $R_{\rm w}$.  
Assuming thermal X-rays as targets, they found $R_{\rm w}\ge 5 \, \rlight$
whereas including also the pulsed component 
tightens this limit to $R_{\rm w}\ge 30 \, \rlight$. Unfortunately,
however, the constraint is not only quite weak in the sense that it does not
extend far into the wind zone, but it is 
also sensitive to the angular dependence of
the pulsar wind: if the conversion from Poynting flux to kinetic energy
affects only the striped wind in the equatorial belt, the gamma-rays would 
be visible only to observers located in this belt, thus invalidating 
the constraint. 

On the other hand, a lower limit on the Lorentz factor of wind follows from the 
fact that induced Compton scattering by the electrons in the wind 
does not appear to influence the radio
pulses significantly. \citet{wilsonrees78} investigated this effect in detail 
and estimated that, for the Crab pulsar wind, 
$\Gamma>10^4$ at a radius of roughly $100\,\rlight$. 

\subsection{Pulses from the wind}

In the striped wind scenario, dissipation of magnetic energy occurs primarily
in the current sheets. If this gives rise to a significant synchrotron
emissivity, the resulting radiation will, as discussed above, appear to be
point-like. Furthermore, provided that $R/\rlight \lesssim \Gamma^2$, where
$R$ is the radius of the radiation source, it will be pulsed at the neutron
star rotation period. This was noticed quite early in the development of 
pulsar theory \citep{michel71,arons79} and has been revived recently
\citep[][see below]{kirketal02}. 
Thus, the striped wind is a possible
site of production of the incoherent, high energy (optical to gamma-ray)
nonthermal radiation observed from numerous pulsars. A priori, there is no
compelling reason to favour this site over the inner or outer gaps in the
magnetosphere. However, in contrast to these theories, the wind model has
the advantage that an analytical description of the magnetic field structure
is available. This removes from its predictions much of the arbitrariness
present in those of the magnetospheric models.

Both the point-like appearance and the 
pulsations of the radiation from the wind are a consequence of
relativistic beaming. 
Assume for simplicity that the striped wind consists
of thin, concentric, spherical, radiating shells. In the equatorial plane,
successive shells, $n$ and $n+1$, are separated by half a wavelength
of the stripes, $\Delta l = \pi\,\rlight$, 
(see figure~\ref{fig:Shell}).  Furthermore,
assume that these shells radiate only after they cross a spherical
surface of radius~$R_0$. The maximum time delay between the arrival
times at a distant observer 
of photons emitted on shell~$n$ is
$\Delta t = \Delta R / c =  ( 1 - \cos\theta)R_0/c$.
For relativistic flows, $\theta \approx 1/\Gamma \ll 1$. Therefore,
\begin{equation}
  \label{eq:PulseWidth} 
  \Delta t \approx \frac{R_0}{2\,\Gamma^2\,c} 
\end{equation}
In order to observe pulses, this time delay should be less than the
time delay between emission of two consecutive shells, $n$ and $n+1$,
crossing $R_0$, given by $\Delta T = \Delta l / c = \pi \, \rlight /
c$. As a consequence, pulses are observed if
\begin{equation}
  \label{eq:Pulsed}
  R_0 \lesssim 2\,\pi\,\Gamma^2\,\rlight 
\end{equation}
\citep[cf.][Eq~(47)]{arons79}.
Using the oblique split monopole solutions, \citet{kirketal02} computed the
pulse profiles expected from the striped wind. In general, two pulses per
period are predicted, as observed in all gamma-ray pulsars. The spacing of
these pulses is uniquely determined by the angles between the rotation axis
and the magnetic axis ($\alpha$) and the rotation axis and the line of sight
($\zeta$).  In the case of the Crab, the pulses are spaced by 0.4 of a period,
consistent with $\alpha=60\degr$ and $\zeta=60\degr$, as independently
suggested by the morphology of the X-ray torus \citep{ngromani04}.  The width
of the main pulse and interpulse observed in the Crab is much larger than
would be expected of a thin current sheet in a wind of Lorentz factor of
$10^5$ or more. Furthermore, there is a significant unpulsed component of
emission.  These properties suggest that at least some of the radiating
electrons diffuse out of the sheet.

As well as pulse profiles, the polarisation of the optical pulses from the
Crab pulsar has motivated detailed comparative studies of the emission
predicted by different magnetospheric models 
\citep{dyksetal04,kaspietal04}. The analogous computations for
the wind model have been performed by \citet{petrikirk05}. 
Extending the oblique, split-monopole solutions, 
these authors model the radial dependence of a current sheet of finite
thickness, including an electron density that peaks in the centre of the
sheet, and falls to a small, but finite value in between them, such that
overall pressure balance is maintained with the varying magnetic field. 
An arbitrary, but small, poloidal component of the 
magnetic field is also 
added, in order to prevent the magnetic field vanishing on
the neutral surface. 

\begin{figure}[htb]
\centering
\includegraphics*[width=9 cm]{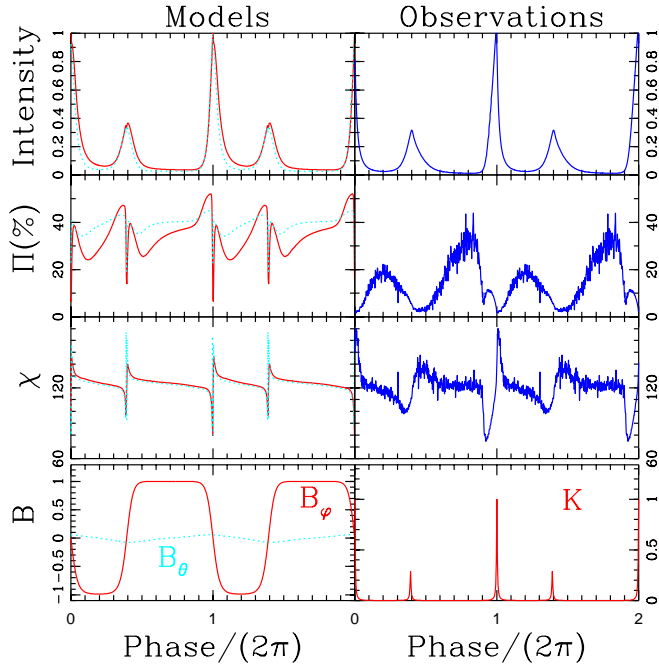}
  \caption{Light curve of intensity, degree of polarisation and 
    position angle of the pulsed synchrotron emission obtained for the
    striped wind model and measurements of these quantities for the
    Crab pulsar.  Models with Lorentz factor~$\Gamma=20$ (solid red)
    and $50$ (dotted cyan) are shown. The bottom panels show the
    dependence on phase of the assumed magnetic field components 
$B_\theta$ and $B_\varphi$ and
    the particle density $K$ in the comoving frame.
    \label{fig:polar}}
\end{figure}

These extensions result in the radial dependences of magnetic field components
and electron density plotted in the lower two panels of
figure~\ref{fig:polar}. The thickness and relative number of electrons in each
sheet is chosen to fit the observed pulse profiles, shown in the top two
panels (model on the left, observations \citep{kanbachetal03} on the right). 
The $B_\theta$ component is chosen to fit the sweep of the linear
polarisation as it enters and leaves the pulse and interpulse (see
the angle of polarisation $\chi$ in figure~\ref{fig:polar}). The fact that 
$B_\theta$ oscillates with the same period as $B_{\varphi}$ implies that the
pulsar wind itself has a small degree of circular polarisation, whose sense is
determined by the sense of rotation of the neutron star, and is, therefore,
the same in both the pulse and interpulse. 

The detailed fits to all
three observed components (intensity, degree of polarisation $\Pi$ and
angle of polarisation $\chi$) are quite good. On the other hand, 
as in the magnetospheric models, a degree of arbitrariness has been 
introduced in order to achieve this. However, one important prediction
of the wind model is independent of these uncertainties. The direction on the
sky of the polarisation vector associated with the d.c.\ component of emission
between the pulses should be determined by $B_\varphi$ alone, i.e., it should lie 
along the projection onto the sky of the rotation axis of the neutron star.
This prediction is testable, because the morphology of the X-ray image of the
nebula enables a symmetry axis to be identified
\citep{ngromani04}. In figure~\ref{fig:polar} this measurement was used to 
orient the model predictions of the angle $\chi$. Thus, the agreement of the 
predicted off-pulse values of $\chi$ with the measured off-pulse polarisation
direction is a strong argument in support of the wind model. 

\section{The termination shock}
\label{terminationshock}

\begin{figure}
\includegraphics[width=10 cm,scale=0.8]{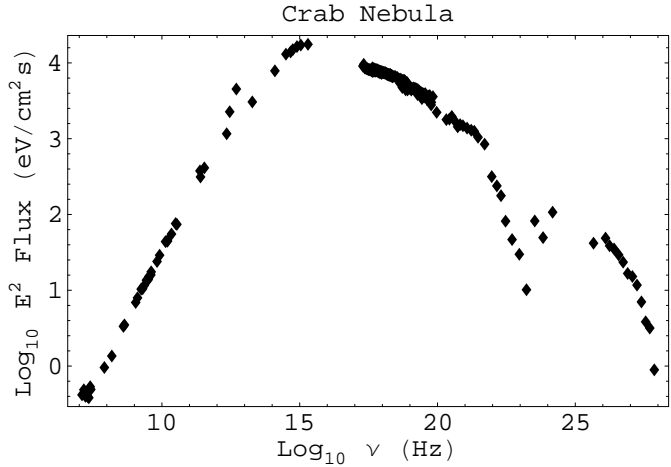}
\caption{%
\label{crabspectr}%
The integrated spectrum of the Crab Nebula.
Radio data are taken from \citet{baarsetal77},
infrared from \citet{greenetal04}, optical from
\citet{veron-cettywoltjer93},
and X-ray through gamma-ray data (EGRET, COMPTEL and BeppoSAX) 
from \citet{kuiperetal01}. 
The TeV data points ($>10^{25}\,$Hz) are from H.E.S.S.\ 
observations, \citep{hesscrab06}.}
\end{figure}

Pulsar wind nebulae (PWNe) 
are observed from the radio to TeV gamma-rays \citep{gaenslerslane06}. 
The spatially integrated spectrum contains information on the 
distribution in energy of the radiating particles that are presumably 
injected at the termination shock. The best observed example, the 
Crab Nebula, is shown in Fig.~\ref{crabspectr}. 
Most of the radiation (from the radio up to $100\,$MeV)
is synchrotron emission, and only the peak in the very-high energy
gamma-ray band is attributed to the inverse Compton scattering of
synchrotron photons on high-energy electrons.
The synchrotron part may be described as power laws 
with spectral breaks around $10^{13}\,$Hz, $10^{15}\,$Hz and 
$100\,$keV. This extremely broad frequency range implies
that the spectrum of relativistic electrons and positrons in the
nebula extends from $\lesssim100\,$MeV to $\sim 1\,$PeV. The
energy density (and pressure) of this relativistic plasma is
dominated by leptons with an energy of around $100\,$GeV.

The spectra of other PWNe are generally similar to that of the Crab.
In the radio band, they are hard: ${\cal
F}_{\nu}\propto \nu^{-\alpha}$, with $\alpha$ between $0$ and $0.3$. 
At high frequencies the spectrum softens, and 
in the X-ray band $\alpha>1$.

 \citet{KC84b} postulated that the wind from the
Crab pulsar has a Lorentz factor $\Gamma_{\rm T}\approx 3\times 10^6$
at the termination shock, where a power-law particle spectrum is
created at energies exceeding $E\sim m_{\rm e}c^2\Gamma_{\rm T}\sim 1\,$TeV.
These electrons and positrons emit from the UV to the gamma-ray
bands. Optically emitting electrons appear in the nebula as a result of
the synchrotron cooling of the 
TeV electrons. The formation of the
power-law spectrum at $E>m_{\rm e}c^2\Gamma_{\rm T}$ was considered by
\citet{hoshino92} and \citet{amatoarons06}. They suggested that the pulsar
wind is loaded by ions. Ion cyclotron waves are then collectively
emitted at the shock front, and positrons and electrons are accelerated by
resonant absorption of these waves. Their simulations of a
relativistic shock in an electron-positron-ion plasma show that
a power-law spectrum of positrons and electrons is formed. 
This model not only accounts for the optical to X-ray
spectrum of the Crab but offers an explanation of the variable
synchrotron emission of the wisps
observed in the vicinity of the termination
shock \citep{scargle69, hesteretal02}.
\citet{gallantarons94} and \citet{spitkovskyarons04} argue that the
wisps arise in regions where reflection of the ions in the
self-consistent magnetic field causes compressions of the
electron-positron plasma and, thus, enhanced emission. The
characteristic variability time-scale of the wisps (a few months) is
determined in this model 
by the Larmor period of ions with Lorentz factors of
a few$\times 10^6$.

The main problem with this model is that it does not account for the
radio to IR emission of the nebula, which is generated by electrons and
positrons with energies between $100\,$MeV 
and $100\,$GeV. The large 
number of 
these electrons and positrons requires a
pair injection rate of $\dot N=10^{40}$ to $10^{41}\,\textrm{s}^{-1}$
\citep{reesgunn74}, implying
$\kappa\sim 10^6$ (see Eq.~\ref{multiplicity}).
The spin-down power of the pulsar $L_{\rm sd}$, when divided by
this pair output, yields a Lorenz-factor of the wind
$\Gamma_{\rm T}=6\times 10^4(10^{40}/\dot N)$, incompatible with the value
of a few$\times 10^6$ required by the ion model.
\citet{KC84b} and \citet{atoyan99} avoided this problem 
by assuming
that the low energy electrons were injected at a very early
stage of the history of the nebula. The synchrotron lifetime of the radio
emitting electrons is significantly longer than
the age of the nebula, so that one cannot exclude 
{\em a priori} that the overall spectrum
depends on history of the nebula. Nevertheless, a convincing explanation of 
this rather convenient event in the past is lacking.  
Another possibility, proposed
by \citet{arons98}, is that the radio and X-ray emitting
particles are injected by different sectors of the termination
shock, X-ray particles being accelerated in the equatorial
belt and the radio ones in the polar region. However, the
apparent continuity of the overall spectrum of the nebula from the
radio to the gamma-ray band favours for a single population of
emitting electrons. 
Moreover, recent observations
of the wisps in the radio band \citep{bietenholzetal01, bietenholzetal04}
suggest that the radio emitting electrons are currently accelerated 
in the same region as those responsible for the optical to
X-ray emission.

The spectral slope of the Crab  is $\alpha=0.3$ in the radio band
\citep{baarsetal77} and  $\alpha=0.72$ in the optical
\citep{veron-cettywoltjer93}, which is compatible with the
assumption that the break at about $10^{13}\,$Hz is due to
the synchrotron burn-off effect in a
magnetic field of $300\,\mu$G, which is close to the equipartition value
\citep{marsdenetal84}. 
The implied energy spectrum of the injected
electrons and positrons has the form $N(E)\propto E^{-1.6}$.
This view is supported by
\citet{gallanttuffs00,gallanttuffs02} who found that the infra-red
spectral index in the central parts of the Crab is close to that
in the radio, and gradually steepens as one moves outwards. 
The high frequency break lies in the ultra-violet band at about a
few$\times 10^{15}\,$Hz, which implies that at $E>E_{\rm br}\sim 1$ TeV
the injection spectrum becomes steeper\footnote{\citet{delzanna06}
argue that the UV break is due to the synchrotron cooling, which
assumes that the magnetic field in the nebula remains a few times
less than the equipartition level. In this case, the $10^{13}\,$Hz
break should be attributed to the break in the injection spectrum
so that $E_{\rm br}\sim 100\,$GeV.}; the spectral slope in the X-ray
band, $\alpha=1.1$, corresponds, when the synchrotron
burn-off effect is taken into account, 
to an injection spectrum of $N(E)\propto E^{-2.2}$ at $E>E_{\rm br}$,
so that the injection spectrum is a broken power-law. Taking into account that
no sign of a low frequency cut-off is observed in the Crab
spectrum down to about $30\,$MHz (at lower frequencies the spectrum
is dominated by the pulsar), one concludes that the injection
spectrum of electrons and positrons extends down to $E_{\rm min}\le 100\,$MeV.

Such an injection spectrum implies a rather unusual acceleration
process. The energy density of the injected plasma, $\int
E\,N(E)\diff E$, is dominated by particles with $E\sim E_{\rm br}$ whereas
most of the particles find themselves at $E\sim E_{\rm min}\ll E_{\rm br}$
so that the acceleration process should somehow transfer most of
the total energy of the system to a handful of particles leaving
for the majority only a small fraction of the energy. This is
not what one would normally expect from a 
shock-associated acceleration process, where
the particle flow is randomised at the shock and only a fraction of
the upstream kinetic energy is converted into the energy of 
a few accelerated
particles. In this case, the temperature $T$ downstream
roughly corresponds to 
the upstream particle kinetic energy
and the high-energy tail of accelerated particles merges, 
at its low energy end, with the
quasi-thermal distribution at $E\sim \textrm{few}\times 
k_{\rm B} T$. For the Crab, this means that the Lorentz factor
of the wind at the termination shock, $\Gamma_{\rm T}$, 
should not exceed a few hundred in
order to deliver the majority of the electrons and positrons 
into the downstream region with $E_{\rm min}\sim 100$ MeV. 
On the other hand, the particle energy
spectrum implies that the energy per electron in the system is
much larger, about $10$ GeV, so there would appear to exist an energy
reservoir in the flow that eventually dissipates in 
a small fraction of the particles.

\citet{gallantetal02}, modifying the original idea of
\citet{hoshino92}, suggested that the wind is loaded by so many 
ions that their kinetic energy dominates the wind energy flux.
At the shock front, the pairs are thermalized with $k_{\rm B}T\sim
m_{\rm e}c^2\Gamma_{\rm T}\leq 100\,$MeV whereas the ions collectively emit
about one half of their energy as cyclotron waves. The radio
emitting electrons and 
positrons are accelerated by these
waves according to the mechanism by \citet{hoshino92}
and \citet{amatoarons06}. 
The problem with this theory, is that the required
injection rate of ions, $\sim L_{\rm sd}/(m_{\rm p}c^2\Gamma_{\rm T})
\sim 10^{39}\,\textrm{s}^{-1}$, vastly exceeds the fiducial Goldreich-Julian elementary
charge loss rate, $\dot N_{\rm GJ}\sim 10^{34}\,\textrm{s}^{-1}$,
obtained by substituting $\kappa=1$ into eq.~(\ref{multiplicity}).
It is difficult to imagine how the electric field at the surface of the star
could extract ions with a density orders of magnitude higher than that required
to screen out this same field. 
On the other hand, one cannot
exclude by observation that pulsars emit the required number of
ions. Their presence in the plasma injected 
into the nebula could, in principle, be revealed by
observations of ultra high energy gamma-rays and/or neutrinos
\citep{amato03,bednarek03,bednarekbartosik03}; the data currently
available on
the Crab Nebula are, however, 
compatible with the
assumption that all the observed emission is generated by
electrons and positrons and no relativistic ions are present
\citep{hesscrab06}.
However, it has recently been proposed that the
high energy emission of the PWN Vela~X 
can be understood
if a significant fraction of the pulsar spin-down power is carried
by relativistic nuclei \citep{horns06}.

Another possible energy reservoir for particle acceleration
is present if most of the pulsar spin-down energy is still stored in
the striped magnetic field when the flow enters the termination
shock \citep{lyubarsky03b,petrilyubarsky07}. 
As discussed in section~\ref{stripedwind}, this would be expected if
the microphysics of the dissipation process proceeds relatively slowly. 
The flow then accelerates only slowly, and the 
Lorentz factor at the termination shock
is compatible with a low $E_{\rm min}$ in the energy distribution of the
accelerated particles. When the flow enters the shock, the
alternating magnetic fields annihilate transferring the energy to
the particles and one can speculate that the radio-to-optical
emission of PWNe is generated by electrons and positrons
accelerated in the course of reconnection of the alternating
magnetic field at the pulsar wind termination shock.
One-dimensional PIC simulations indeed show that the alternating
fields easily annihilate at the shock \citep{lyubarsky05,petrilyubarsky07}, 
but do not
show evidence of a nonthermal distribution. This may be attributed to a highly
idealised one-dimensional field structure in the simulations. In
the real case, reconnection is expected to occur at \hbox{X-points},
and particles gain energy according to how close they approach such a
point. Higher dimensional studies of the reconnection process
in a relativistic electron-positron plasma without a striped field
\citep{romanovalovelace92, zenitanihoshino01, zenitanihoshino05a,
zenitanihoshino05b, larrabee03} demonstrate efficient
acceleration of non-thermal particles.  However, even though the
obtained spectra can be roughly described by a power-law with an
exponential cutoff, there is as yet no evidence of 
a hard power-law spectrum over the wide energy range observed in PWNe.

The formation of the high-energy tail in the spectrum at $E>E_{\rm br}$, 
on the other hand, can be attributed to the first-order Fermi acceleration
mechanism. 
In the absence of strong cross-field diffusion, 
this mechanism does not 
operate at a perpendicular
shock (where the magnetic field is perpendicular to the shock normal), 
because particles are prevented from diffusing back
upstream by the fact that their guiding centres must follow the
field lines, which run parallel to the shock front. 
Because the perpendicular field component is compressed in the 
proper frame by a large factor, relativistic shocks almost always fall into 
this category \citep{begelmankirk90}.   
However, in reality, some degree of cross field transport must occur, and the 
question of whether or not the first-order mechanism operates at a 
relativistic shock hinges on the relative magnitudes of the ordered magnetic
field and the stochastic component that produces the cross-field transport.
Simulations of acceleration in prescribed stochastic fields indeed show that
acceleration is less effective for stronger ordered fields 
\citep{ostrowskibednarz02}, and corroborating 
evidence is beginning to accumulate from 3D, relativistic,  
particle-in-cell simulations \citep{spitkovsky05}.
On the other hand, in the pulsar case, 
annihilation of the ordered magnetic field in the wind (or in 
the shock) may leave a highly turbulent,
small-scale magnetic field, whose chaotic
component significantly exceeds the regular field
\citep{revilleetal06}. Particle
scattering off these strong inhomogeneities may then
allow the first-order Fermi mechanism to operate. It is
interesting that the first-order Fermi mechanism
operating at an ultra-relativistic shock yields, 
in the case of isotropic diffusion of the accelerated particles,
an energy distribution $E^{-2.2}$ \citep{bednarzostrowski98,
kirketal00, achterbergetal01} --- exactly the value required 
to explain the X-ray spectrum of the Crab.

\section{The nebula}
\label{nebula}

The physical conditions in pulsar winds, though difficult 
to determine directly,
can be inferred from the observed
properties of PWNe, which are simply the 
bubbles inflated by the wind in the surrounding medium. 
It is the wind plasma that fills the nebula and produces the
observed nonthermal electromagnetic emission. The typical size of
the PWN is a few parsecs. Specifically, the Crab Nebula is an ellipsoid
of projected dimension $2\,$pc$\,\times\,3\,$pc.

The nebula is confined by its nonrelativistic surroundings, and
the termination shock of the pulsar wind is located at a
radius defined by the condition that the confining pressure
balances the momentum flux of the wind. In the case of the Crab Nebula, the
shock radius was estimated by \citet{reesgunn74} to be $0.1\,$pc,  in
excellent agreement with the radius of the apparent central hole
in the nebula brightness distribution \citep{scargle69}. At the
shock front, the wind energy is released into the relativistic
particles responsible for the observed radiation. Rapidly moving
wisps and variable knots were found in this region
\citep{scargle69,hesteretal95,hesteretal02}, confirming it as the site of 
energy injection into the nebula.

Early spherically symmetrical MHD models of the Crab Nebula
\citep{reesgunn74, KC84a, KC84b, emmeringchevalier87} seem to
describe its main properties perfectly well. The observed
brightness and the spectral index distributions are generally
consistent with the assumption that the relativistic particles are
injected by the pulsar in the centre of the nebula and then expend
their energy on synchrotron emission and $p\diff V$ work 
\citep{KC84b,amato00}. The size of the nebula decreases with observed
frequency because the higher the energy of a particle, the
faster it loses energy by synchrotron emission. The synchrotron
life-time of the radio emitting electrons is larger than the age
of the nebula. Therefore, they fill the whole nebula. The life-time
of the optically emitting electrons is comparable or somewhat shorter 
than the age of the nebula and, therefore, the optical image is smaller
than the radio one. The X-ray emitting electrons lose their
energy in only a few years. Therefore, X-rays are emitted only
close to the pulsar. This makes the X-rays an
exceptional tool for the study of the interaction of the pulsar
wind with the nebula. Whereas the radio and optical images are
dominated by the outer parts of the nebula and distorted by
interaction with the surroundings, the X-rays trace the freshly
injected plasma and reveal the physical processes in the very heart
of the nebula.

The basic conclusion of these early models was that the pulsar wind must
be particle dominated. 
The magnetic field in the lab.\ frame 
increases by a factor of roughly 
three across a relativistic 
shock front. In the subsonic postshock flow, the plasma density remains
approximately
constant, so that conservation of the magnetic flux within a
radially expanding magnetic loop implies linear growth of the
field strength with radius. In order for the magnetic pressure
at the outer boundary of the nebula not to exceed the plasma
pressure, the magnetisation of the wind just upstream of the
termination shock should be at most $\sigma\sim3\times
10^{-3}$ \citep{KC84a,emmeringchevalier87}. Such a low value of
$\sigma$ is puzzling, because the pulsar
wind is launched as a Poynting dominated flow with $\sigma\gg 1$, and,
as discussed in sections~\ref{aligned} and \ref{stripedwind}, 
it is not easy to invent a realistic mechanism
to reduce $\sigma$ to the required level.

On the other hand, $\sigma$ cannot be significantly lower
than the above value, because magnetic stresses play an important role 
in shaping the nebula: the observed elongation
is convincingly explained as the result of pinching 
by an azimuthal magnetic field
\citep{begelmanli92, vanderswaluw03}. The discovery of the jet-torus
structure confirmed that the nebula is elongated along the pulsar
spin axis so that the magnetic field in the nebula should indeed be
wrapped around the major axis of the ellipsoid.

The jet-torus structure was revealed already in earlier
X-ray observations \citep{aschenbachbrinkmann75,brinkmannetal85,hesteretal95} 
and was clearly
resolved by Chandra \citep{weisskopfetal00, hesteretal02,mori04}. 
Similar structures have been found
in other PWNe \citep[for a review, see][]{gaenslerslane06}, suggesting that
this is a generic phenomenon. The inner boundary of the X-ray torus
of the Crab 
coincides with the position of the standing shock predicted by
early spherically symmetric models, but the morphology
of the inner nebula forces one to abandon 
the assumption of spherical symmetry. 
The observed structure suggests that the pulsar wind itself is
highly anisotropic, with most of the energy being transported in the
equatorial belt. Encouragingly, simple (split-monopole) 
models of the pulsar wind indeed have the property that the energy flux 
reaches a maximum at the equator (see section~\ref{stripedwind}).
However, the origin of the polar jet is not so evident.

The jet in the Crab Nebula, as well as the jets in other PWNe, appears to
originate from the pulsar and propagate along the rotation axis. This seems
to indicate that they are a part of the pulsar wind, possibly
collimated by magnetic hoop stresses. 
However, as discussed in section~\ref{aligned}, 
this mechanism does not work in 
ultra-relativistic flows. Moreover,
such a jet would presumably be ultrarelativistic, whereas the observed
jets certainly are not, as follows both from direct measurements of the
proper motion in the jets of the Crab and Vela 
\citep{hesteretal02, pavlov03}, and from the fact that
both the jet and counter-jet are visible.
\citet{lyub02} pointed out that magnetic collimation,
though ineffective in the pulsar wind, 
could be responsible for the formation of the jets
{\em beyond} the termination shock. 
In the equatorial
belt, which carries most of the energy, annihilation of the
striped field (see section~\ref{stripedwind}) ensures that the
residual magnetic field is low and does not affect the dynamics of
the post-shock plasma. However, the magnetisation of the high
latitude flow remains significant. This naturally results in the
separation of the post-shock flow into an equatorial disk and a
magnetically collimated polar outflow. In this model, the observed
jet arises as a result of the axial compression of the shocked
plasma. 

In the split monopole model of the pulsar wind, the angular
distribution of the total energy flux $f_{\rm tot}$  
can be written as
\eqb
    f_{\rm tot}&=&\frac{f_0}{R^2}(\sin^2\theta+1/\sigma_0)
\label{eflux}
\eqe
\citep{michel73,bogovalov99},
where $R$ and $\theta$ are the usual spherical coordinates, $f_0$
and $\sigma_0\gg 1$ constants. The first term in parentheses
represents the Poynting flux, whereas the second one accounts for
the small initial contribution of particles. As discussed
 above, a significant part of the Poynting flux
is transferred by variable fields, and can be 
converted into plasma energy relatively easily. 
But the total energy per
particle is conserved along the streamlines and, as they are
radial, the angular distribution (\ref{eflux}) remains unchanged.
\citet{lyub02} showed that in such a strongly anisotropic wind, the
termination shock is highly aspherical, forming a cusp on the
axis of the flow. 
Therefore, the jet could appear to originate from the pulsar simply
because the termination shock lies much closer to the pulsar
in the polar regions than in the equatorial belt.
By neglecting magnetic stresses in the vicinity of the shock,
\citet{bogovalkhang02} independently found 
that the subsonic region lies much closer to the pulsar
in the polar region than in the equatorial plane. 
However, the formation of a jet on the axis 
depends on the additional effect of magnetic collimation
\citep{khangbogoval03}.

To test this qualitative picture
\citet{komissarov03, komissarov04} performed axisymmetric
relativistic MHD simulations of a flow produced by an anisotropic
pulsar wind within a slowly expanding cavity of cold heavy gas.
They adopted eq.~(\ref{eflux}) for the angular distribution of the
total energy in the wind and assumed that all waves decayed either
in the wind or at the termination shock so that the postshock flow
is completely determined by the total energy flux, $f_{\rm tot}$
and the mean field, $B$, in the wind.  As the postshock MHD
parameters are independent of where exactly the waves decay, one
can assume for simplicity that all the wave energy has already
been converted into the flow kinetic energy
on entering the shock, $\rho\Gamma
v=f_{\rm tot}-cB^2/4\pi$. For an ultra-relativistic flow, $\Gamma\gg
1$, the post-shock plasma is relativistically hot and the dynamics
of the downstream flow depends only on the total energy flux and
magnetisation, i.e., it does not depend on 
$\rho$ and $\Gamma$ separately, but only on their product.  
Therefore, the flow is determined by the
two functions, $f_{\rm tot}(\theta)$ and $B(\theta)$.

The distribution of the mean field in the pulsar wind is not known
but certainly goes to zero on the equator of a
flow driven by an obliquely rotating, centred dipole. Moreover the mean
field vanishes on the axis of the flow, because an unphysical
singular current would otherwise be required. Taking
into account that the field is frozen into a radial flow and,
therefore, the radial dependence is given by
eq.~(\ref{asymptoticsol}), \citet{komissarov03, komissarov04} 
chose a simple model for the mean field:
\begin{equation}
   B=\sqrt{\frac{4\pi f_0}c}\,\frac{\xi}{R}\,
   \sin\theta\left(1-\frac{2\theta}{\pi}\right)
\label{Bwind}
\end{equation}
where the free parameter $\xi\le 1$ controls the magnetisation of
the wind. The ratio of the energy transported by the mean
electromagnetic field to the energy transported by the particles
is $\sigma=0.1\xi^2$.

\begin{figure}
\includegraphics[bb=56 150 282 385,clip=true,width=\textwidth]{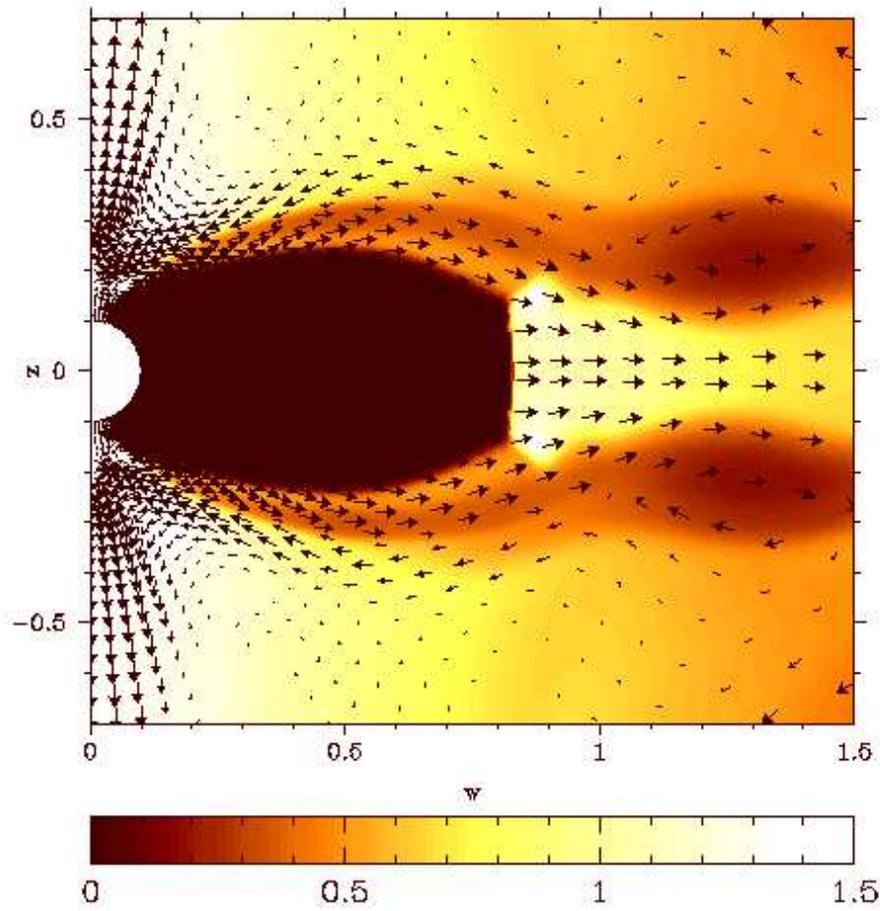}
\caption{%
\label{fig:flow}%
The flow
around the termination shock \citep{komissarov03}. The flow
velocity is shown by arrows, the plasma energy density by colour. 
}
\end{figure}

The results of these simulations are shown in Fig.~\ref{fig:flow}.
The termination shock is highly aspherical
being significantly closer to the pulsar in the polar zone than in
the equatorial plane. Most of the downstream flow is initially
confined to the equatorial plane. The magnetic hoop stresses stop
the outflow in the surface layers of this equatorial disk and
redirect it into magnetically confined polar jets, which, therefore,
are formed outside of the termination shock. Velocities both in the disk
and in the jet were found to be about $0.5c$, close to those
inferred from observation \citep{hesteretal02,pavlov03,delaney06}.

\begin{figure}
\includegraphics[bb=82 146 288 390,clip=true,width=\textwidth]{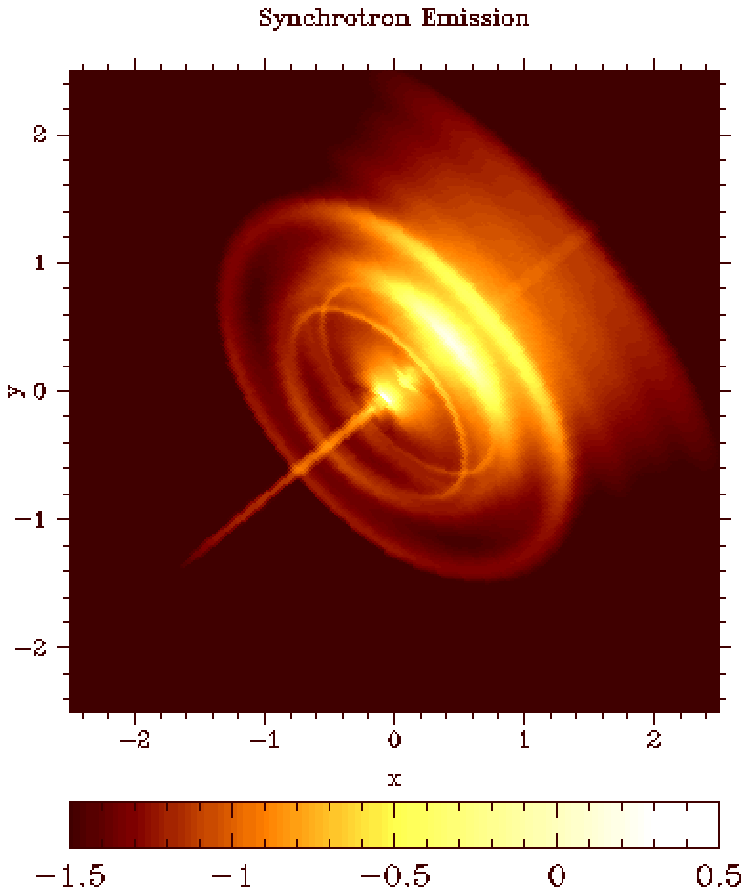}
\caption{%
\label{fig:image}%
The simulated image of the synchrotron emission of the Crab Nebula
\citep{komissarov03}. }
\end{figure}

Images of the nebula in synchrotron emission
were also simulated, taking
into account the relativistic beaming effect and particle energy
losses (Fig.~\ref{fig:image}). These images resemble those of
the Crab and other PWNe obtained by Chandra. They exhibit both a
system of rings, giving the impression of an equatorial disk-like
or even toroidal structure, and well-collimated polar jets, which
give the illusion of originating directly from the pulsar.

The simulated images also reveal a bright central source.
At high latitudes, plasma enters the shock
highly obliquely, and, therefore, 
the post-shock velocity remains
high, $\sim0.9c$ and is not deflected far from its upstream
radial direction. 
As discussed in section~\ref{pointlike}, this results in 
an almost point-like feature close to the pulsar. However, it
is not connected with emission from the Crab pulsar itself, but
may be identified with the bright knot discovered by
\citet{hesteretal95} and located $0.65\,$arcsec 
to the southeast.

\citet{delzanna04} investigated 
the sensitivity of the morphology of the nebula
to the angular dependence of the wind magnetisation. They
simulated the evolution of the pulsar wind nebula within the expanding
supernova ejecta adopting Eq.~(\ref{eflux}) for the total energy
flux, but parameterising the angular dependence of the mean magnetic field 
as $B\propto\sin\theta\tanh[b(\pi/2-\theta)]$ (cf.\ 
eq.~(\ref{Bwind})). The effect on the flow of the 
width of the low magnetisation
region in the equatorial belt (associated with the striped wind)
was investigated by varying $b$. They found that collimation
occurs at any $b$ 
but that the overall picture of the flow changes completely if
there is no belt of low magnetisation, i.e. when
$b\to\infty$.

In this case, the magnetic hoop stresses in the equatorial plane
completely suppress the radial flow after a few
termination shock radii, diverting the plasma towards the axis.
A part of the flow then enters a polar outflow
whereas another part is directed towards the pulsar forming a
large-scale vortex. The polar outflow is quite wide (the radius
is comparable with the equatorial radius of the termination shock)
and is formed rather far from the pulsar. The outflow starts at
a distance larger than the equatorial radius of the termination
shock, whereas, below the base of the outflow, the plasma moves
towards the pulsar. A wide outflow starting far from the
pulsar was also found by \citet{bogovalov05} who simulated the
nebula formed by a pulsar wind with $B\propto\sin\theta$. This
picture does not match the observed structure of the
nebula,  where the jet is very narrow and appears to start
close to the pulsar.

The clearly observed polar-equator dichotomy arises only if
the magnetisation is low in the equatorial belt of the pulsar wind;
according to \citet{delzanna04}, $b=10$ is enough. Only in this
case is the equatorial outflow not significantly 
affected by the
magnetic stresses, and is able to extend deep into the
nebula, whilst the high latitude wind is collimated
into a narrow polar jet. In this sense, the jet-torus structure 
provides evidence in favour of magnetic dissipation in
the striped wind. One can speculate that the smaller the angle
between the magnetic and rotation axes of the pulsar, the narrower
the striped zone and, therefore, the larger is the 
fraction of the flow that is collimated into a polar jet.

The MHD model thus captures many properties of the Crab Nebula quite well and,
in spite of some quantitative differences with the observational data, is
probably basically correct.  By incorporating more physics into the model, one
can hope to achieve even better agreement with the observations and infer the
parameters of the pulsar wind in different systems.  Thus,
\citet{bucciantini05} computed polarisation maps of the nebula, and
\citet{delzanna06} investigated the effects of synchrotron cooling on the
images and spectral properties of PWNe.  \citet{bucciantini06} addressed the
development of the Kelvin-Helmholtz instability in the disk outflow and the
resulting small-scale modulation of the synchrotron radiation. Global
variations of the flow are also clearly visible in simulations;
\citet{bogovalov05} has reported quasiperiodic oscillations on timescales from
years to dozens of years, but the origin of these variations is still unclear.

The interaction of a PWN with the surrounding gas is 
another phenomenon that can be studied using relativistic MHD simulations,
and the results are a valuable tool in 
the interpretation of the wide variety shown by X-ray images. 
Thus, the evolution of the nebula within the expanding supernova
ejecta, including interaction with the reverse shock, was
simulated by \citet{vanderswaluw01} in a spherically symmetric
model, and
\citet{bucciantini04} studied the Rayleigh-Taylor instability at
the interface between an expanding PWN and its surrounding
supernova remnant.  
\citet{bucciantini05a} presented axisymmetric
simulations of pulsar wind bow-shock nebulae arising around
pulsars, that have already emerged from the progenitor supernova
remnant.

These detailed results are strongly dependent on the angular structure of the
pulsar wind.  Recent progress in simulating the obliquely rotating dipole
magnetosphere and wind \citep{spitkovsky06}, suggests that a more realistic
model of the angular dependence of the pulsar wind might soon be available to
replace parameterisations such as eqs.~(\ref{eflux}) and (\ref{Bwind}).  Then
one might even hope to constrain the angle between the rotation and magnetic
axes of the pulsar from the observed structure of the inner nebula.

All PWNe models discussed so far are either spherically or
axially symmetric. However, in MHD flows with
non-negligible magnetic stresses, 3D effects could come into 
play because the underlying symmetric configurations could be
unstable.  Inadequacy of the axisymmetric picture may be suggested
by the fact that all models of the Crab Nebula require a very low 
overall magnetisation $\sigma\approx 0.01$ --- 
if the magnetisation were larger the nebula would be distorted by
the pinch effect beyond observational limits. 
However, the 
ideas on magnetic dissipation discussed in
section~\ref{stripedwind} are capable of removing only the oscillating part
of the magnetic field in the equatorial belt. This apparent 
problem can be alleviated
if, as suggested by \citet{begelman98}, the kink instability destroys the 
concentric field structure in the nebula.
In the axisymmetric case, magnetic
loops in the expanding flow are forced to expand and perform work 
against the magnetic tension. The kink instability 
allows the loops to come apart and one can expect that in 3D, 
the hoop stress would be less effective than suggested by 
axisymmetric simulations. These stresses cannot disappear altogether, 
because
they are responsible for driving the kink instability 
itself. Nevertheless, this effect could have
important implications for the inferred magnetisation of the
pulsar wind.

\section{Summary}

The suggestions of \citet{piddington57} that the relativistic particles and
magnetic fields in PWNe originate in a central stellar object, as well as that
of \citet{pacini67} that the ultimate energy source lies in the rotational
energy of the neutron star, have both stood the test of time and the scrutiny
of increasingly high resolution observations at all accessible wavelengths.
We also now know that the pulsar wind has differing equatorial and polar
components, as suggested by \citet{aschenbachbrinkmann75}. The Crab Nebula, in
particular, can be quite well described by an axisymmetric, relativistic MHD
model. This establishes without much doubt that the energy is injected into
the nebula at the
wind termination shock, with most of it being concentrated 
into an equatorial belt in a particle dominated form, and that the jets 
consist of shocked plasma collimated by azimuthal fields, rather than a polar
wind. 

Consensus is more difficult to achieve in
the case of the wind. That it is highly relativistic on entering the
termination shock, with bulk Lorentz factor
$\Gamma_{\rm T}>100$, seems secure. This lower limit, which 
applies only if the 
flow remains magnetically dominated up to the shock front, is still
controversial; other models generally place $\Gamma_{\rm T}$ between
$10^4$ and $10^6$. That the equatorial flow is born as a 
\lq\lq striped wind\rq\rq\ 
\citep{michel71,coroniti90} seems likely. That it is almost perfectly radial
and accelerates at least to the fast magnetosonic point is also not
controversial. 

Perhaps the most puzzling open question is the matter content of 
the wind. Detailed models of pair creation in the magnetosphere produce far
too few leptons to explain all of the nebular emission. The ion content is
also puzzling: observations of the wisps or, alternatively, 
of the integrated spectrum, could
be nicely explained if ions contributed to the energy flux, but  
they are even harder to extract from the pulsar.

The much discussed $\sigma$--problem also remains an open question,
especially for the Crab Nebula.
Although two solutions work for the equatorial part of the wind (dissipation
of the stripes either in the wind or at the termination shock), neither of
these works in the polar wind region. The only possibilities here seem to be
either that the equatorial belt is initially very broad (i.e., a highly
oblique rotator) or that kink instabilities in the outer nebula ultimately
release the magnetic tension. 

The particles responsible for the nonthermal emission are 
almost certainly accelerated at the termination shock front. However,
which mechanisms are responsible is controversial. It seems that at least two 
different ones must be operating, one at low energies, ($100\,$MeV to 
$1\,$TeV) which may be either resonant absorption of 
coherently emitted ion cyclotron waves, or the annihilation of magnetic flux
in the shock front, and one at high energy ($>1\,$TeV) which is most likely
the first order Fermi mechanism. However, none of these mechanisms can be
regarded as fully worked out from a theoretical point of view. 

Progress, as always, will flow from more and improved observations. 
For the nebula, 
X-ray and optical images continue to be of crucial importance, and 
maps in the TeV band are just starting to appear 
\citep{hess_msh15-52_05}. For the wind, the observational signatures will be
point-like, and might be pulsed. High time-resolution optical polarimetry of
pulsar emission will have important input on the question of the location of
the pulse-forming region. Finally, 
the large number of new gamma-ray pulsars that will
be discovered by GLAST is also highly likely to lead to the extinction of some
theories of high energy pulsed emission, as well as providing much more
accurate spectra for comparison with models of PWNe. 
\begin{acknowledgement}
This work was supported by a grant from the G.I.F.\ the German-Israeli
Foundation for Scientific Research and Development
\end{acknowledgement}
%
% BibTeX users please use
%\bibliographystyle{aa}
%\bibliography{ms.bib}

%

\printindex
\end{document}